\documentclass[10pt, conference, letterpaper]{IEEEtran}
\IEEEoverridecommandlockouts
\usepackage[
  letterpaper, 
  left=0.65in,   
  right=0.65in,  
  top=0.75in,    
  bottom=1.05in   
]{geometry}
\usepackage[T1]{fontenc}
\usepackage{times} 
\usepackage[scaled]{helvet} 
\usepackage[hyphens]{url}
\usepackage[hidelinks, bookmarks=false]{hyperref}
\usepackage{cite}
\usepackage{amsmath,amssymb,amsfonts}
\usepackage{algorithmic}
\usepackage{graphicx}
\usepackage{comment}
\usepackage{threeparttable}
\usepackage{booktabs}
\usepackage{textcomp}
\usepackage{colortbl, xcolor}
\definecolor{bestgreen}{RGB}{198,239,206}
\definecolor{bestgreen}{RGB}{198,239,206}
\usepackage{array} 
\usepackage{multirow} 
\usepackage{colortbl, xcolor}   
\usepackage{makecell}           
\definecolor{bestgreen}{RGB}{198,239,206}
\usepackage{hyperref}
\usepackage{algorithm}
\usepackage{algorithmic}
\usepackage{mathtools}
\usepackage{amsthm}
\newtheorem{definition}{Definition}
\newtheorem{proposition}{Proposition}

\usepackage{bm}
\usepackage{tikz}
\usetikzlibrary{arrows.meta, positioning, fit, backgrounds, decorations.pathreplacing}
\usepackage{xcolor}
\def\BibTeX{{\rm B\kern-.05em{\sc i\kern-.025em b}\kern-.08em
    T\kern-.1667em\lower.7ex\hbox{E}\kern-.125emX}}

\newcommand{\R}{\mathbb{R}}
\newcommand{\C}{\mathbb{C}}
\newcommand{\hilbert}{\mathcal{H}}
\newcommand{\fourier}[1]{\hat{#1}}
\newcommand{\inner}[2]{\langle #1, #2 \rangle}
\newcommand{\norm}[1]{\left\| #1 \right\|}
\newcommand{\abs}[1]{\left| #1 \right|}
\def\BibTeX{{\rm B\kern-.05em{\sc i\kern-.025em b}\kern-.08em
    T\kern-.1667em\lower.7ex\hbox{E}\kern-.125emX}}

\begin{document}
\bstctlcite{IEEEexample:BSTcontrol}
\title{Continuous Orthogonal Mode Decomposition: Haptic Signal Prediction in Tactile Internet}

\author{\IEEEauthorblockN{Mohammad Ali Vahedifar, Mojtaba Nazari, Qi Zhang}
\IEEEauthorblockA{DIGIT and Department of Electrical and Computer Engineering, Aarhus University, Denmark}
\thanks{This research was supported by the TOAST project, funded by the European Union’s Horizon Europe research and innovation program under the Marie Skłodowska-Curie Actions Doctoral Network (Grant Agreement No. 101073465), the Danish Council for Independent Research project eTouch (Grant No. 1127- 00339B), and NordForsk Nordic University Cooperation on Edge Intelligence (Grant No. 168043). Authors' e-mails: \{av, qz\}@ece.au.dk.

\textbf{Link to the Code: \href{https://github.com/Ali-Vahedifar/Continuous-Orthogonal-Mode-Decomposition}{GitHub Code.}}}
}

\maketitle

\begin{abstract}
The Tactile Internet demands sub-millisecond latency and ultra-high reliability, as even slight latency or packet loss can destabilize haptic control. To address this, we propose the Mode-Domain Architecture (MDA), a bilateral predictive neural network architecture designed to restore missing signals on both the human and robot sides. Unlike conventional models that implicitly extract features from raw data, MDA employs a novel Continuous-Orthogonal Mode Decomposition framework. By integrating an orthogonality constraint, we overcome the pervasive issue of ``mode overlapping’’ found in state-of-the-art decomposition methods. Experimental results demonstrate that this structured feature extraction achieves high prediction accuracies of 98.6\% (human) and 97.3\% (robot). Furthermore, the model achieves ultra-low inference latency of 0.065 ms, significantly outperforming existing benchmarks and meeting the stringent real-time requirements of haptic teleoperation.
\end{abstract}

\begin{IEEEkeywords}
Mode Decomposition, Haptic Signal Prediction, Tactile Internet, Ultra-low latency and High Reliability.
\end{IEEEkeywords}

\section{Introduction}
\label{sec:intro}

The Tactile Internet (TI) is designed to enable a human operator to interact with a remote robot by receiving real-time force feedback, allowing intuitive execution of complex tasks~\cite{Antonakoglou}. However, TI faces two critical challenges.  First, communication delay between the human operator’s command and the robot’s (teleoperator’s) response results in perceived loss of transparency and system instability. Second, packet loss or missing signals can cause the signals from both sides to become uncorrelated, thereby failing to accurately preserve the subtle statistical properties of the motion on the robot side. The TI research community has shown growing interest in integrating general Machine Learning (ML) methods for signal prediction; such models offer promising avenues toward more stable and practical remote control~\cite{salvato2022predicting, GeorgeGlobecom2025, LeFo,GPSFV}. The main limitation of these approaches is that they operate directly on raw time-series data. Although such models can minimize statistical prediction errors, they often fail to distinguish the underlying physical mechanisms that drive the signal. 

From the perspective of semantic communication, a haptic signal is not a monolithic stream but a superposition of distinct physical events, such as the operator’s low-frequency intent, high-frequency environmental textures and sudden contact transients. Predicting force ($F$), velocity ($V$), or position ($P$) on raw data forces the model to learn these conflicting dynamics simultaneously, which can lead to poor generalization on complex tasks. This motivates us to focus primarily on prediction in the ``mode’’ domain. This approach posits that haptic signals should be treated not as single patched objects but as composites of distinguishable components. This enables the predictive model to learn the specific behavior of each infrastructure component independently.

The main contributions of this paper are as follows:\\
\textbf{i. Continuous-Orthogonal Mode Decomposition (C-OMD):} 
We enforce orthogonality in the continuous mode space. This formulation enables zero-mode interference.\\
\textbf{ii. Mode-Domain Architecture (MDA):} 
We introduce a tailored architecture that operates in the decomposed mode domain for bilateral signal prediction in TI. MDA achieves significantly higher accuracy and \textbf{8-30$\times$} inference speedup.
\section{Related Works}\label{RW}

Signal decomposition techniques are adaptive tools for breaking down complex waveforms into their constituent components. Among these approaches, Variational Mode Decomposition (VMD)~\cite{DragomiretskiyVMD} defines modes as amplitude-modulated, frequency-modulated (AM–FM). The VMD requires the number of modes to be specified a priori, which is often a challenge in practice. Based on VMD, extensions such as Variational Mode Extraction (VME)~\cite{NazariVME} and Successive Variational Mode Decomposition (SVMD)~\cite{SVMDNAZARI2020107610} aim to alleviate parameter sensitivity through iterative mode extraction and residual analysis.  However, VME extracts only the desired mode rather than performing a full decomposition, and SVMD replaces VMD's parallel structure with successive computations that depend on previously extracted modes. This sequential dependency reduces compatibility with GPU-based implementations, where parallelism is critical to overcoming computational bottlenecks.

These approaches rely on solving approximated augmented Lagrangian formulations. Due to these approximations, complete mode separation is not theoretically guaranteed, and mode interference may still occur. Moreover, existing methods either require the number of modes to be specified in advance or sacrifice parallelizability. These limitations motivate the development of a new framework that preserves the parallel structure. We recast the problem of selecting the number of modes as an empirical grid search and provide theoretical guarantees for complete mode separation. In this work, we propose C-OMD. The key idea is to enforce orthogonality directly in the mode space, thereby enabling principled, theoretically grounded separation.

\section{Preliminaries}
\label{sec:preliminaries}

\begin{definition}[Intrinsic Mode]
An intrinsic mode is modeled as an amplitude-modulated and frequency-modulated (AM-FM) signal:
\begin{equation}
    m_k(t) = A_k(t)\cos(\phi_k(t)),
    \label{eq:am_fm_mode}
\end{equation}
where the instantaneous amplitude $A_k(t)$ and the instantaneous frequency $\nu_k(t) = \phi_k'(t)$ vary much more slowly than the phase $\phi_k(t)$ itself.
\end{definition}
While \eqref{eq:am_fm_mode} provides the physical representation of a mode, this highly non-linear formulation is not directly amenable to variational optimization or the enforcement of global orthogonality constraints. However, as noted in VMD \cite{DragomiretskiyVMD}, a smoothly varying AM-FM signal naturally exhibits a compact frequency spectrum centered around a mean frequency $\omega_k$, the amplitude-weighted average of $\nu_k(t)$. C-OMD therefore bounds the modes through the following bandwidth characterization.
\begin{definition}[Mode]
\label{def:mode}
A function $m_k \in H^2(\R)$ is a \emph{mode} with center frequency
$\omega_k \in [0,\infty)$ if its spectral energy is concentrated near $\omega_k$,
as measured by the normalized spectral variance:
\begin{equation}
    B_k^2 := \frac{
    \displaystyle\int_0^{\infty}
    (\omega - \omega_k)^2\,\abs{\fourier{m}_{k,+}}^2\,d\mu
    }{
    \displaystyle\int_0^{\infty}
    \abs{\fourier{m}_{k,+}}^2\,d\mu
    },
    \label{eq:mode_bandwidth}
\end{equation}
where $\fourier{m}_{k,+}(\omega)$ is the Fourier transform of the analytic
signal of $m_k$.
\end{definition}
We work in the Hilbert space $L^2(\R)$ of square-integrable functions
equipped with the inner product:
\begin{equation}
    \inner{u}{v}  =  \int_{-\infty}^{\infty} u(t)\,\overline{v(t)}\,dt
    \label{eq:inner_product}
\end{equation}
where $\overline{v(t)}$ denotes the complex conjugate of $v(t)$, and the induced norm $\norm{u}_{L^2} = \inner{u}{u}^{1/2}$. Modes are further required to lie in the Sobolev space $H^2(\R) = \{u \in L^2(\R) : u'' \in L^2(\R)\}$, which keeps Fourier-domain integrals involving $\omega^2$ finite and lets Parseval's isometry apply. We write $\fourier{m}(\omega)=\int m(t)e^{-j\omega t}dt$ and $d\mu \triangleq d\omega/2\pi$, so that $\inner{m}{n} = \int \fourier{m}\,\overline{\fourier{n}}\,d\mu$.

\begin{definition}[Orthogonal Mode Decomposition]
A signal $f \in L^2(\R)$ admits an \emph{orthogonal mode decomposition of
order $K$}, with $K$ finite and fixed a priori, if there exists a collection
$\{m_k\}_{k=1}^{K}$ of intrinsic modes such that
\begin{equation}\label{eq:mode_reconstruction}
    \sum_{k=1}^{K} m_k(t) = f(t), \quad
    G_{ij} \triangleq \inner{m_i}{m_j} = 0, \quad \forall\, i \neq j.
\end{equation}
\end{definition}

\begin{definition}[Analytic Signal]
For a real-valued signal $m(t) \in L^2(\R)$, the analytic signal is:
\begin{equation}
    m_+(t)  =  m(t) + j\,\hilbert[m(t)],
    \label{eq:analytic_signal}
\end{equation}
where $\hilbert[\cdot]$ is the Hilbert Transform~\cite{Hahn1996}. Its Fourier transform is supported only on non-negative frequencies:
$\fourier{m}_+(\omega) = 2\fourier{m}(\omega)$ for $\omega \geq 0$ and
$\fourier{m}_+(\omega) = 0$ for $\omega < 0$.
\end{definition}

\section{Continuous-Orthogonal Mode Decomposition}
\label{sec:continuous_vmd}
The central contribution of C-OMD is to augment the orthogonality constraint between modes. The problem seeks modes $\{m_k\}_{k=1}^{K} \subset H^2(\R)$ and
center frequencies $\{\omega_k\}_{k=1}^{K} \subset [0,\infty)$ that solve:
\begin{equation}
\begin{aligned}
    &\min_{\{m_k\},\,\{\omega_k\}}  \sum_{k=1}^{K}
    \norm{\partial_t\!\left[
    m_{k,+}(t)\, e^{-j\omega_k t}
    \right]}_{L^2}^{2} \\
    &\text{s.t.} \quad (\mathrm{i})  \quad  \sum_{k=1}^{K} m_k(t) = f(t)
    \qquad\quad  \text{(reconstruction)} \\
    &\qquad\;\; (\mathrm{ii})  \quad G_{ij} = 0,
       \quad \forall\, i \neq j
    \qquad\text{(orthogonality)}
\end{aligned}
\label{eq:omd_problem}
\end{equation}
where $m_{k,+}(t) = m_k(t) + j\,\hilbert[m_k(t)]$ denotes the analytic signal. Constraint~$(\ref{eq:omd_problem}.\mathrm{i})$ forces the modes to account for the whole signal, and constraint~$(\ref{eq:omd_problem}.\mathrm{ii})$ stops any two of them from carrying the same information. We handle the pair by introducing a Lagrange multiplier $\lambda(t) \in L^2(\R)$ with quadratic penalty $\alpha > 0$ for reconstruction, together with a symmetric multiplier matrix $\bm{\Gamma} = [\gamma_{ij}] \in \R^{K \times K}$, $\gamma_{ii} = 0$, and penalty $\beta > 0$ for orthogonality. The augmented Lagrangian reads:
\begin{equation}
\begin{aligned}
    \mathcal{L}\bigl(\lambda,\bm{\Gamma}\bigr)
     = \sum_{k=1}^{K} \norm{\partial_t\!\left[
    m_{k,+}(t)\, e^{-j\omega_k t}
    \right]}_{L^2}^{2}
     +  \alpha \big\|f - \sum_{k} m_k\big\|_{L^2}^{2}
     \\+  \inner{\lambda}{f - \sum_{k} m_k}
     + \beta \sum_{i \neq j} \abs{G_{ij}}^2
     + \sum_{i \neq j} \gamma_{ij}\,G_{ij}
\end{aligned}
\label{eq:augmented_lagrangian}
\end{equation}
The penalty $\alpha\norm{f - \sum m_k}^2$ converges at finite weight, matching a Bayesian prior with noise variance $\propto 1/\alpha$, while $\lambda(t)$ enforces reconstruction exactly in the limit. The penalty $\beta\sum_{i \neq j}|G_{ij}|^2$ suppresses inter-mode correlation. Note that $\alpha$ weights squared spectral amplitudes and $\beta$ weights squared inner products, so the two carry different physical scales and must be tuned independently.

\subsection{Fourier-Domain Reformulation}
\label{subsec:fourier_domain}

By Parseval's isometry, the Lagrangian~\eqref{eq:augmented_lagrangian}
can be evaluated entirely in the Fourier domain:
\begin{equation}
\begin{split}
    \sum_{k=1}^{K}\!\int_{0}^{\infty}\!\!(\omega - \omega_k)^2
    \abs{\fourier{m}_{k,+}}^2 d\mu
    \;+\;\alpha\!\int_{-\infty}^{\infty}\!\!
    \Bigl|\fourier{f}- \sum_{k}\fourier{m}_k\Bigr|^2 d\mu \\
    +\int_{-\infty}^{\infty}\!\!
    \fourier{\lambda}\,\overline{\Bigl(\fourier{f} - \sum_{k}\fourier{m}_k\Bigr)}\, d\mu
    \;+\sum_{i \neq j} \Bigl( \beta \abs{G_{ij}}^2 + \gamma_{ij} G_{ij} \Bigr).
    \label{eq:ortho_fourier}
\end{split}
\end{equation}
\textbf{Mode Update:} Taking the first variation of~\eqref{eq:augmented_lagrangian} with respect to
$\overline{\fourier{m}_k(\omega)}$ and setting it to zero yields the mode update.
For non-negative frequencies $\omega \geq 0$:
\begin{equation}
    \fourier{m}_{k}^{n+1}(\omega) =
    \frac{\mathcal{A}}{1 + 2(\omega - \omega_k)^2/\alpha},
    \label{eq:mode_update_omd}
\end{equation}
\begin{equation}
\begin{aligned}
    \mathcal{A}\hspace{-1mm}:=\hspace{-1mm}\fourier{f}(\omega) \hspace{-1mm}-\hspace{-1mm} \sum_{i \neq k} \hspace{-1mm}\fourier{m}_i(\omega)
    \hspace{-1mm}+\hspace{-1mm} \frac{\fourier{\lambda}(\omega)}{2\alpha}
    \hspace{-1mm}-\hspace{-1mm} \frac{1}{\alpha}\hspace{-1mm}\sum_{j \neq k}
    \hspace{-.5mm}\Bigl(\beta\,\inner{m_k^n}{m_j^n} \hspace{-1mm}+\hspace{-1mm} \frac{\gamma_{kj}}{2}\Bigr)
    \fourier{m}_j(\omega).
\end{aligned}
\label{eq:numerator_A}
\end{equation}
\textbf{Center Frequency Update:} This is unchanged from VMD~\cite{DragomiretskiyVMD}, being independent of the
orthogonality constraint:
\begin{equation}
    \omega_k^{n+1}  = 
    \frac{
    \displaystyle\int_{0}^{\infty} \omega\,\abs{\fourier{m}_k^{n+1}}^2\,d\mu
    }{
    \displaystyle\int_{0}^{\infty}
    \abs{\fourier{m}_k^{n+1}}^2\,d\mu
    },
    \label{eq:center_freq_update}
\end{equation}
which places $\omega_k$ at the center of gravity of the mode's power spectrum.

\textbf{Dual Ascent:} The Lagrange multipliers are updated by dual (gradient) ascent:
\begin{align}
    \lambda^{n+1}(t) & = 
    \lambda^n(t) + \tau_\lambda\Bigl(f(t) - \sum_{k}m_k^{n+1}(t)\Bigr),
    \label{eq:lambda_update} \\[4pt]
    \gamma_{ij}^{n+1} & = 
    \gamma_{ij}^n + \tau_\gamma\,\inner{m_i^{n+1}}{m_j^{n+1}},
    \quad i \neq j,
    \label{eq:gamma_update}
\end{align}
where $\tau_\lambda, \tau_\gamma > 0$ are the dual step sizes.

\subsection{Limitations of Penalty-Only Orthogonality Enforcement}
\label{subsec:penalty_limitations}

Eq.~\eqref{eq:numerator_A} handles orthogonality through the correction term $-\frac{1}{\alpha}\sum_{j \neq k}(\beta\inner{m_k^n}{m_j^n} + \gamma_{kj}/2)\fourier{m}_j(\omega)$ in the numerator. This steers the modes toward mutual orthogonality, but only indirectly: the correction comes from the inner products of the \emph{previous} iterate, and the Wiener filter then reshapes the spectral coefficients to satisfy bandwidth, reconstruction, and orthogonality at once, three objectives weighted against each other by $\alpha$ and $\beta$. Two consequences follow.\\
\textbf{i. Slow convergence to orthogonality.} The quadratic penalty $\beta\sum_{i\neq j}|G_{ij}|^2$ punishes non-orthogonality but never projects onto the orthogonal manifold. At finite $\beta$ the modes are only approximately orthogonal, and high precision demands $\beta \to \infty$, which destabilizes the bandwidth objective.\\
\textbf{ii. Conflict with the Wiener filter.} The denominator $1 + 2(\omega - \omega_k)^2/\alpha$ enforces spectral compactness around $\omega_k$. The numerator correction subtracts components correlated with other modes, and the bandwidth filter then re-weights that subtraction. A single Wiener-type update therefore buys a trade-off, not both exact compactness and exact orthogonality.

We address this by augmenting the ADMM (Alternating Direction Method of Multipliers) iteration with an explicit \emph{orthogonal projection step} after the Wiener update. The scheme is alternating projection: the Wiener filter enforces bandwidth and reconstruction, and a Newton--Schulz step then projects the current estimates onto the nearest orthogonal system. Each step corrects the violation left by the other, and the pair alternates until neither residual improves. The projection runs on the Gram matrix of the mode system, introduced next.

\subsection{Newton--Schulz Iteration for Orthogonality Enforcement}
\label{subsec:ns_functional}
The Newton--Schulz iteration orthogonalizes a system of functions by operating on the $K \times K$ Gram matrix while linearly combining the functions themselves. Given $K$ candidate real-valued modes $\{m_1(t), \ldots, m_K(t)\} \subset L^2(\R)$, the Gram matrix $\mathbf{G} \in \R^{K \times K}$ is defined by:
\begin{equation}
    G_{ij} = \int_{-\infty}^{\infty} \! m_i(t)\,m_j(t)\,dt
    = 2\,\Re\!\int_{0}^{\infty} \!
      \fourier{m}_i\,\overline{\fourier{m}_j}\,d\mu,
    \label{eq:gram_matrix}
\end{equation}
where the second equality follows from Parseval's theorem together with the
Hermitian symmetry of real signals, so that $\mathbf{G}$ is assembled directly
from the one-sided spectra. The modes are orthogonal if and only if
$\mathbf{G}$ is diagonal.

Let $\{m_k^{\text{in}}(t)\}_{k=1}^K$ be the modes returned by the Wiener
update~\eqref{eq:mode_update_omd} and normalize (superscript in stands for initial):
\begin{equation}
    m_k^{(0)}(t) =
    \frac{m_k^{\text{in}}(t)}{\bigl(\sum_{j}\norm{m_j^{\text{in}}}_{L^2}^2\bigr)^{1/2}},
    \label{eq:ns_normalize}
\end{equation}
which places the eigenvalues of $\mathbf{G}^{(0)}$ inside the convergence
region. Compute $\mathbf{G}^{(0)}$ from the \emph{normalized} modes
via~\eqref{eq:gram_matrix} and iterate, for $\ell = 0,1,\ldots,L-1$:
\begin{equation}
    \mathbf{C}^{(\ell)} =
    \tfrac{3}{2}\mathbf{I}_K - \tfrac{1}{2}\mathbf{G}^{(\ell)},
    \label{eq:ns_coefficient}
\end{equation}
\begin{equation}
    m_k^{(\ell+1)}(t) =
    \sum_{j=1}^{K} C_{kj}^{(\ell)}\, m_j^{(\ell)}(t),
    \label{eq:ns_function_update}
\end{equation}
\begin{equation}
    G_{ij}^{(\ell+1)} =
    \inner{m_i^{(\ell+1)}}{m_j^{(\ell+1)}}.
    \label{eq:ns_gram_update}
\end{equation}
Since \eqref{eq:ns_normalize} discards the energy required by constraint~$(\ref{eq:omd_problem}.\mathrm{i})$, we close the projection with the per-mode rescaling $m_k^{\perp} = (\norm{m_k^{\text{in}}}_{L^2}/\norm{m_k^{(L)}}_{L^2})\,m_k^{(L)}$, a positive diagonal map that scales $G_{ij}^{(L)}$ by $D_{ii}D_{jj}$ and so restores the Wiener-scale energies without disturbing orthogonality.
\begin{figure}[t]
    \centering
    \includegraphics[width=\linewidth]{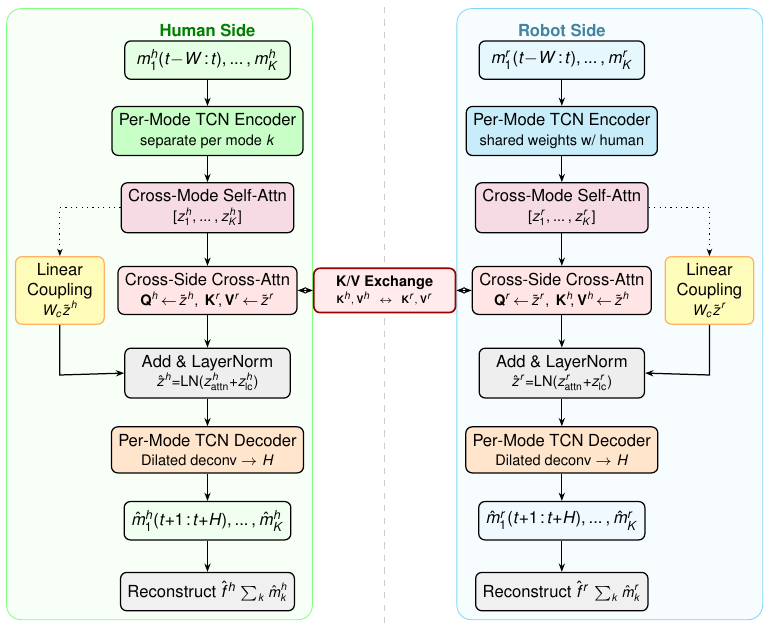}
    \caption{Mode-Domain Architecture. C-OMD decomposes each haptic signal into $K$ orthogonal modes processed by per-mode Temporal Convolutional Network (TCN) encoders and Cross-Mode Self-Attention within each side. Cross-Side Cross-Attention (main path) lets the human latent attend to robot keys/values and vice versa; a parallel Linear Coupling branch (dotted skip) provides a stable residual $W_c\tilde{z}$ added before LayerNorm, ensuring gradient flow even when attention weights are noisy. Fused latents $\hat{z}^h,\hat{z}^r$ are decoded and summed to reconstruct the predicted signals.}
    \label{fig:architecture}
\end{figure}
\begin{figure}[t]
    \centering
    \includegraphics[width=\linewidth]{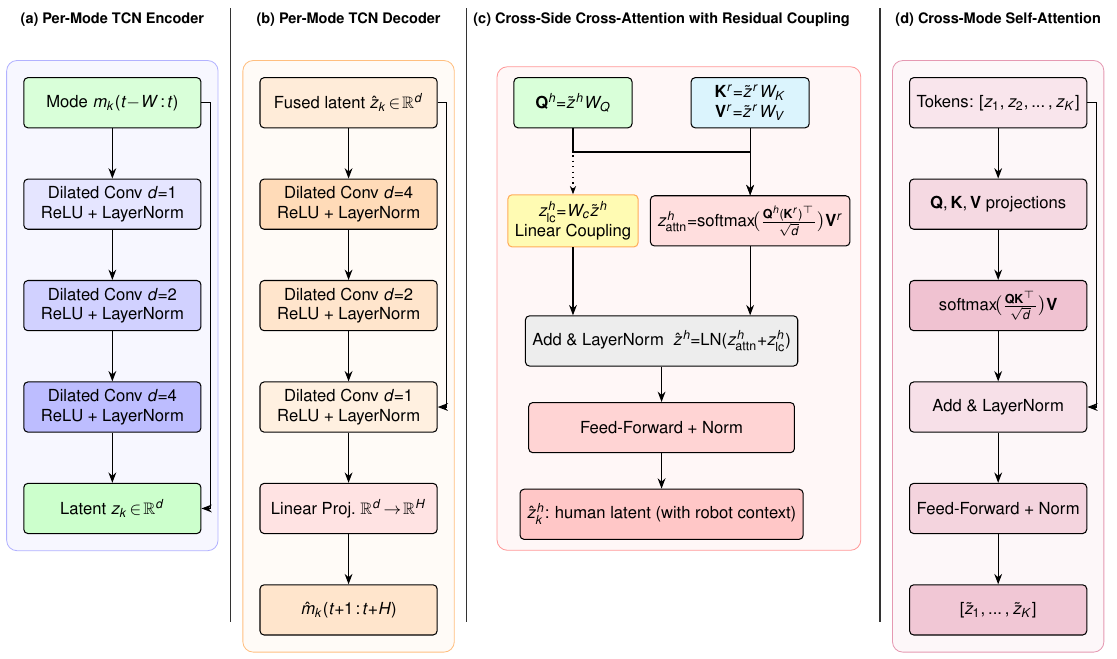}
    \caption{\textbf{(a)}~Per-mode TCN encoder: dilations $d{=}1,2,4$ with residual skip.
  \textbf{(b)}~Per-mode TCN decoder: inverted dilations $d{=}4,2,1$ with
  linear projection $\mathbb{R}^d\!\to\!\mathbb{R}^H$.
  \textbf{(c)}~Cross-Side Cross-Attention with residual coupling detail:
  $z_\text{attn}$ and $z_\text{lc}{=}W_c\tilde{z}$ are summed and
  layer-normalized.
  \textbf{(d)}~Cross-Mode Self-Attention over $K$ mode latents.}
    \label{fig:architecture2}
\end{figure}
\subsection{Fourier-Domain Realization of the Projection}
\label{subsec:per_freq_ortho}
The functional iteration of Section~\ref{subsec:ns_functional} is realized entirely in the Fourier domain, which keeps every transform out of the inner loop. At each frequency $\omega$, collect the spectral amplitudes of the $K$ modes into
\begin{equation}
     \mathbf{v}(\omega) = [\fourier{m}_1(\omega), \ldots,
     \fourier{m}_K(\omega)]^\top \in \C^{K} .
    \label{eq:freq_vector}
\end{equation}
The composite Newton--Schulz transform $\mathbf{P} = \prod_{\ell=L-1}^{0}\mathbf{C}^{(\ell)} \in \R^{K\times K}$ is formed once per ADMM iteration from the frequency-integrated Gram matrix
\begin{equation}
    \mathbf{G} = 2\,\Re\!\int_{0}^{\infty}
    \mathbf{v}\,\mathbf{v}^H\,d\mu,
    \label{eq:freq_gram}
\end{equation}
and then applied pointwise in $\omega$:
\begin{equation}
    \mathbf{v}^{\perp}(\omega)  =
    \mathbf{P}\,\mathbf{v}(\omega), \qquad \forall\,\omega \geq 0 .
    \label{eq:per_freq_projection}
\end{equation}
Since $\mathbf{G}$ is real, so is $\mathbf{P}$, which preserves Hermitian symmetry and returns real-valued modes. Carrying no frequency dependence, $\mathbf{P}$ makes \eqref{eq:per_freq_projection} exactly \eqref{eq:ns_function_update} written spectrally, at $O(K^2)$ per frequency bin and with no FFT: \eqref{eq:gram_matrix} builds $\mathbf{G}$ straight from the one-sided spectra.
\begin{figure*}[t]
    \centering
    \includegraphics[width=\linewidth]{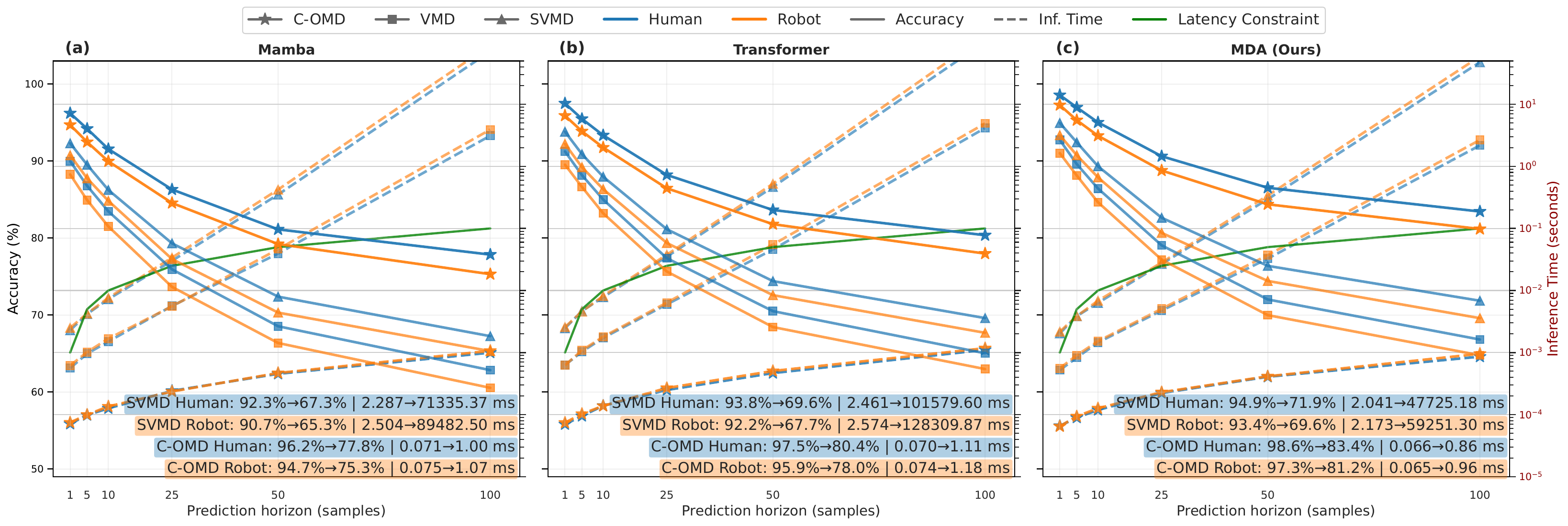}
      \caption{%
    (a,b,c) Prediction accuracy (left axis, solid lines) and inference time
    (right log-axis, dashed lines) vs.\ prediction horizon
    $H \in \{1,5,10,25,50,100\}$ samples for VMD~(\protect$\square$),
    SVMD~(\protect$\triangle$), and C-OMD~(\protect$\bigstar$) across three architectures.
    The \textcolor{green!60!black}{green} line marks the TI
    latency constraint $\Delta t = H\,\text{ms}$ at $f_s{=}1\,$kHz.
  }
  \label{fig:acc_vs_horizon}
\end{figure*}

\section{Experiments \& Performance Evaluation}\label{Exp}

\textbf{Dataset.} We use real-world haptic data traces captured with a Novint Falcon haptic device~\cite{rodriguez_guevara_2025_14924062}. The dataset records 3-DoF position, velocity, and force measurements at $f_s{=}1000\,\text{Hz}$ from both the human operator and the teleoperated robot. All reported results are averaged over 10 independent runs. Our data split is chronological, not random: each task trace is split into contiguous training, validation, and test segments (70/10/20\%) \emph{before} any sliding window is generated, and windows are drawn only within a segment. Consecutive segments are separated by a guard gap of $W{+}H$ samples, so no window ever straddles a boundary and no test window shares a sample with a training window.\\
\textbf{Training.} MDA is trained by a four-component loss:
\begin{align}
    \mathcal{J} = \mathcal{J}_{\mathrm{pred}}
                + \lambda_1 \mathcal{J}_{\mathrm{recon}}
                + \lambda_2 \mathcal{J}_{\mathrm{orth}}
                + \lambda_3 \mathcal{J}_{\mathrm{rel}},
    \label{eq:loss}\\
    \mathcal{J}_{\mathrm{pred}} = \tfrac{1}{KH}\textstyle\sum_{k,h}
        \bigl\|\hat{m}_k(t{+}h) - m_k(t{+}h)\bigr\|_2^2,\nonumber\\
    \mathcal{J}_{\mathrm{recon}} = \bigl\|\hat{f} - f\bigr\|_2^2, \nonumber \qquad
    \mathcal{J}_{\mathrm{orth}} = \textstyle\sum_{i\neq j} |G_{ij}|^2,\\
    \mathcal{J}_{\mathrm{rel}} = \tfrac{1}{KH}\textstyle\sum_{k,h}
    \frac{|\hat{m}_k(t{+}h) - m_k(t{+}h)|}
         {\max(|m_k(t{+}h)|,\,\delta)} \nonumber.           
\end{align}
$\mathcal{J}_{\mathrm{pred}}$ is the primary per-mode MSE prediction loss over window $H$. $\mathcal{J}_{\mathrm{recon}}$ penalizes reconstruction error after summing predicted modes back to the full signal, ensuring the modes remain jointly consistent. In $\mathcal{J}_{\mathrm{orth}}$ the $\mathbf{G}_{ij}{=}\langle m_i, m_j\rangle$, enforces explicit orthogonality of the decomposed modes. $\mathcal{J}_{\mathrm{rel}}$ is a relative error term with floor $\delta{=}0.01$ to avoid division by near-zero values during low-force regions near contact transitions. This term prevents the model from ignoring perceptually important small-force events that pure MSE training tends to underweight. Loss weights are set to $\lambda_1{=}0.1$, $\lambda_2{=}0.01$, and $\lambda_3{=}0.05$.

We report accuracy as $\mathrm{Acc} = 100\bigl(1 - \norm{\hat{f}-f}_2/\norm{f}_2\bigr)\%$, computed on the reconstructed signal over the full test horizon and averaged over the three signal types. Accuracy and inference time are evaluated across prediction horizons $H \in \{1,5,10,25,50,100\}$. At training time, the MDA decoder observes a mixture of ground-truth and model-predicted mode values as its input sequence~\cite{lamb2016professor}. Let $\epsilon_e = 1 - e/E$ denote the \emph{teacher-forcing ratio} at epoch $e$ out of $E$ total epochs. With probability $\epsilon_e$, the decoder receives the true data sequence; otherwise, it receives the model's own prediction from the previous step.\\
\textbf{Inference.} At inference time, when a haptic packet is lost or delayed beyond the predefined deadline, the model enters an \emph{autoregressive restoration} mode: the predicted signal output from the previous step is concatenated with the available history and fed back as the encoder input for the next step, without any ground-truth observations.\\
\textbf{Architecture.} We used Mamba~\cite{mamba}, Transformer~\cite{vaswani2017attention}, and MDA (An overview of the proposed architecture is presented in Fig.~\ref{fig:architecture}, while detailed views of each section are provided in Fig.~\ref{fig:architecture2}) as our Neural Network (NN) architectures. All models are trained with Adam ($\beta_1{=}0.9$, $\beta_2{=}0.999$), cosine annealing with a 5-epoch linear warm-up, weight decay $10^{-4}$, and gradient clipping at norm~1.0. Learning rates are $3{\times}10^{-4}$ for Mamba and Transformer and $5{\times}10^{-4}$ for MDA. NNs trained for 300 epochs on a single NVIDIA RTX Ada~6000.

\section{Result and Discussion}
\begin{figure*}[t]
    \centering
    \includegraphics[width=\linewidth]{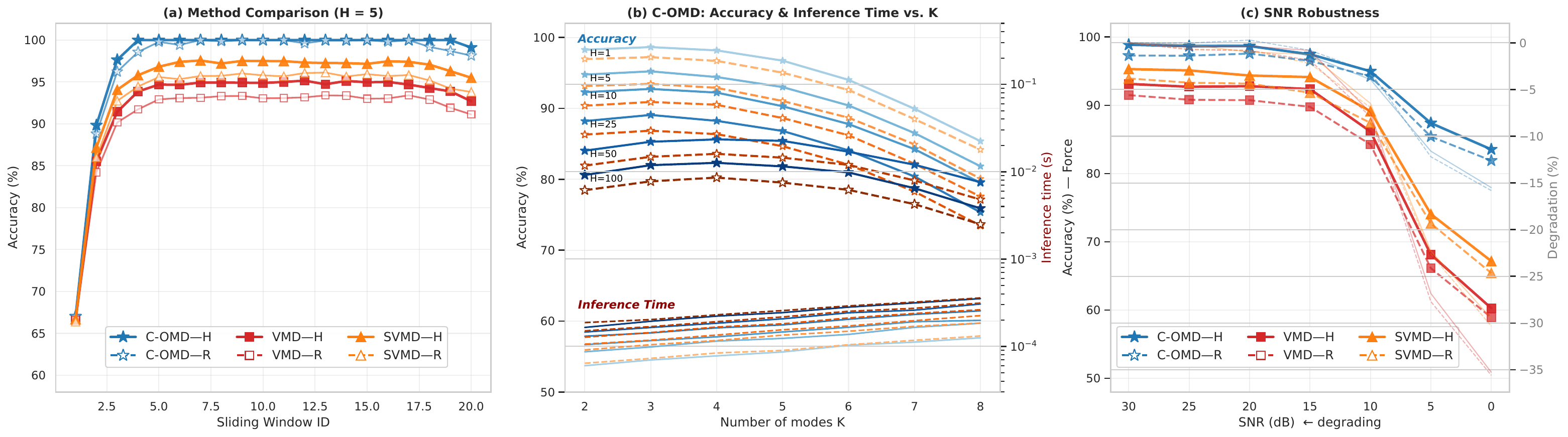}
  \caption{
    \textbf{(a)}~Prediction accuracy over sliding windows across three architectures at horizon $H{=}5$.
    \textbf{(b)} Prediction accuracy (left axis, solid) and inference time (right log-axis, dashed) vs.\ number of modes $K \in \{2,\ldots,8\}$ for C-OMD with the MDA architecture, evaluated at $H \in \{1,5,10,25,50,100\}$. \textbf{(c)}~SNR robustness on the Force signal: accuracy (\%, left axis) and relative degradation (\%, right axis, faint) as channel SNR degrades from $30\,\text{dB}$ to $0\,\text{dB}$.
  }
  \label{fig:combined_results}
\end{figure*}
\begin{table*}[tbp]
\caption{
\textbf{Total GFLOPs}: decomposition + model floating-point operations per
  forward pass ($K{=}3$, $H{=}100$).
  \textbf{Inf.\ (ms)}: inference time at $H{=}1$ per signal type and side.
  \textbf{Acc.\ (\%)}: mean\,$\pm$\,std over 5-fold at $H\in\{1,10,100\}$.
  \cellcolor{bestgreen}Green\,=\,best; \underline{underline}\,=\,second-best.%
}

\label{tab:combined}
\centering
\setlength{\tabcolsep}{1pt}
\begin{threeparttable}
\footnotesize{
\begin{tabular}{@{}ll
                c           
                cccccc      
                ccc         
                ccc@{}}     
\toprule
& & &
  \multicolumn{3}{c}{\textbf{Inf.\ (ms) — Human}}
& \multicolumn{3}{c}{\textbf{Inf.\ (ms) — Robot}}
& \multicolumn{3}{c}{\textbf{Accuracy (\%) — Human}}
& \multicolumn{3}{c}{\textbf{Accuracy (\%) — Robot}} \\
\cmidrule(lr){4-6}\cmidrule(lr){7-9}\cmidrule(lr){10-12}\cmidrule(lr){13-15}
\textbf{Arch.} & \textbf{Decomp.}
  & \textbf{GFLOPs}
  & \textbf{F} & \textbf{V} & \textbf{P}
  & \textbf{F} & \textbf{V} & \textbf{P}
  & $H{=}1$ & $H{=}10$ & $H{=}100$
  & $H{=}1$ & $H{=}10$ & $H{=}100$ \\
\midrule
\multirow{3}{*}{Mamba}
  & VMD
    & 0.28
    & 0.49 & 0.49 & 0.48 & 0.51 & 0.51 & 0.50
    & $89.8{\scriptstyle\pm0.8}$ & $81.4{\scriptstyle\pm1.1}$ & $61.2{\scriptstyle\pm1.9}$
    & $88.1{\scriptstyle\pm0.9}$ & $79.7{\scriptstyle\pm1.2}$ & $59.0{\scriptstyle\pm2.1}$ \\
  & SVMD\tnote{\dag}
    & 1.85
    & 1.95 & 1.96 & 1.93 & 2.02 & 2.04 & 2.01
    & $92.3{\scriptstyle\pm0.7}$ & $84.6{\scriptstyle\pm1.0}$ & $65.8{\scriptstyle\pm1.7}$
    & $90.7{\scriptstyle\pm0.8}$ & $83.0{\scriptstyle\pm1.1}$ & $63.5{\scriptstyle\pm1.9}$ \\
  & C-OMD
    & \cellcolor{bestgreen}0.13
    & \cellcolor{bestgreen}\textbf{0.06} & \cellcolor{bestgreen}\textbf{0.06} & \cellcolor{bestgreen}\textbf{0.06}
    & \cellcolor{bestgreen}\textbf{0.06} & \cellcolor{bestgreen}\textbf{0.07} & \cellcolor{bestgreen}\textbf{0.06}
    & $96.2{\scriptstyle\pm0.5}$ & $90.1{\scriptstyle\pm0.8}$ & $76.4{\scriptstyle\pm1.4}$
    & $94.7{\scriptstyle\pm0.6}$ & $88.5{\scriptstyle\pm0.9}$ & $74.1{\scriptstyle\pm1.5}$ \\
\midrule
\multirow{3}{*}{\makecell[l]{Trans.\\+Fourier}}
  & VMD
    & 0.32
    & 0.51 & 0.51 & 0.50 & 0.53 & 0.54 & 0.52
    & $91.2{\scriptstyle\pm0.7}$ & $83.5{\scriptstyle\pm1.0}$ & $63.5{\scriptstyle\pm1.8}$
    & $89.6{\scriptstyle\pm0.8}$ & $81.8{\scriptstyle\pm1.1}$ & $61.2{\scriptstyle\pm2.0}$ \\
  & SVMD\tnote{\dag}
    & 1.98
    & 2.04 & 2.05 & 2.02 & 2.12 & 2.14 & 2.10
    & $93.8{\scriptstyle\pm0.6}$ & $86.4{\scriptstyle\pm0.9}$ & $68.1{\scriptstyle\pm1.6}$
    & $92.2{\scriptstyle\pm0.7}$ & $84.8{\scriptstyle\pm1.0}$ & $65.9{\scriptstyle\pm1.8}$ \\
  & C-OMD
    & \cellcolor{bestgreen}0.17
    & \cellcolor{bestgreen}\textbf{0.06} & \cellcolor{bestgreen}\textbf{0.06} & \cellcolor{bestgreen}\textbf{0.06}
    & \cellcolor{bestgreen}\textbf{0.07} & \cellcolor{bestgreen}\textbf{0.07} & \cellcolor{bestgreen}\textbf{0.07}
    & $\underline{97.4{\scriptstyle\pm0.4}}$ & $\underline{91.8{\scriptstyle\pm0.7}}$ & $\underline{79.2{\scriptstyle\pm1.3}}$
    & $\underline{95.9{\scriptstyle\pm0.5}}$ & $\underline{90.1{\scriptstyle\pm0.8}}$ & $\underline{77.0{\scriptstyle\pm1.4}}$ \\
\midrule
\multirow{3}{*}{\textbf{MDA}}
  & VMD
    & 0.23
    & 0.44 & 0.45 & 0.44 & 0.46 & 0.47 & 0.46
    & $92.6{\scriptstyle\pm0.7}$ & $84.9{\scriptstyle\pm1.0}$ & $65.3{\scriptstyle\pm1.7}$
    & $91.0{\scriptstyle\pm0.8}$ & $83.2{\scriptstyle\pm1.1}$ & $63.1{\scriptstyle\pm1.9}$ \\
  & SVMD\tnote{\dag}
    & 1.73
    & 1.76 & 1.78 & 1.75 & 1.84 & 1.85 & 1.82
    & $94.9{\scriptstyle\pm0.6}$ & $87.8{\scriptstyle\pm0.9}$ & $70.2{\scriptstyle\pm1.5}$
    & $93.4{\scriptstyle\pm0.7}$ & $86.1{\scriptstyle\pm1.0}$ & $68.0{\scriptstyle\pm1.7}$ \\
  & \textbf{C-OMD}
    & \cellcolor{bestgreen}\textbf{0.09}
    & \cellcolor{bestgreen}\textbf{0.06} & \cellcolor{bestgreen}\textbf{0.06} & \cellcolor{bestgreen}\textbf{0.06}
    & \cellcolor{bestgreen}\textbf{0.06} & \cellcolor{bestgreen}\textbf{0.06} & \cellcolor{bestgreen}\textbf{0.06}
    & \cellcolor{bestgreen}$\mathbf{98.6}{\scriptstyle\pm0.3}$
    & \cellcolor{bestgreen}$\mathbf{93.4}{\scriptstyle\pm0.6}$
    & \cellcolor{bestgreen}$\mathbf{82.4}{\scriptstyle\pm1.2}$
    & \cellcolor{bestgreen}$\mathbf{97.3}{\scriptstyle\pm0.4}$
    & \cellcolor{bestgreen}$\mathbf{91.8}{\scriptstyle\pm0.7}$
    & \cellcolor{bestgreen}$\mathbf{80.1}{\scriptstyle\pm1.3}$ \\
\bottomrule
\end{tabular}%
}
\end{threeparttable}
\end{table*}
Accuracy and inference time evaluation across various window sizes are depicted in Fig.~\ref{fig:acc_vs_horizon}. The strongest configuration, C-OMD+MDA, achieves $98.6\%$ (Human) and $97.3\%$ (Robot) in $H{=}1$, gracefully degrading to $82.4\%$ and $80.1\%$ in $H{=}100$, while VMD with the same backbone as MDA collapses to $65.3\%$ and $63.1\%$, a degradation rate more than twice as steep.

The inference time comparison is equally decisive for real-time deployment. The parallel orthogonalization of C-OMD is ${\approx}8{\times}$ faster than VMD and ${\approx}30{\times}$ faster than SVMD, with C-OMD+MDA completing a full forward pass in under $0.06\,\text{ms}$ at $H{=}1$, well within the $1\,\text{ms}$ TI budget even at $H{=}100$. SVMD, despite offering a modest accuracy gain over VMD (+2 -- +5\%), pays a prohibitive latency penalty from its sequential mode extraction, consistently exceeding $1.7\,\text{ms}$ at $H{=}1$ and trending toward or beyond the latency constraint at longer windows (Fig.~\ref{fig:acc_vs_horizon}, green line). This makes SVMD architecturally incompatible with strict haptic deadlines.

Each point in Fig.~\ref{fig:acc_vs_horizon}. (a, b, c) Represents the average accuracy across all window slides for a specific $H$. For example, when $H= 5$, the point reflects the average accuracy over 20 window slides for the next 100-sample prediction. To examine this in greater detail, see Fig.~\ref{fig:combined_results}. (a) illustrates the changes in accuracy across each sliding window for the 100-sample prediction with $H= 5$. The architecture is fixed to MDA while the decomposition method varies: C-OMD achieves a mean accuracy of \textbf{98.6\%} (Human) and \textbf{97.3\%} (Robot), outperforming VME by \textbf{+3.7} and VMD by \textbf{+6.0}, with a consistently faster cold-start convergence (within 3--4 windows). This confirms that \emph{explicit spectral orthogonality is the dominant performance driver}: eliminating cross-mode leakage stabilizes the latent representation from the first window onward.

The most striking finding from Fig.~\ref{fig:combined_results}.(b) is that
\emph{more modes do not mean better performance}, a result that
Challenges the conventional assumption that richer signal decompositions always benefit from prediction. For VMD and SVMD (Results provided in the GitHub repository due to the page limit), accuracy peaks at $K{=}4$--$6$ depending on the window and then degrades because additional quasi-orthogonal modes introduce redundant spectral content that the MDA cannot distinguish from noise. C-OMD breaks this pattern entirely: its explicit orthogonality constraint forces each mode to carry unique, non-overlapping information, so the optimal $K$ collapses to just $K{=}3$ across \emph{all} horizons $H \in \{1,\ldots,100\}$, achieving $98.6\%$ at $H{=}1$ and $82.4\%$ at $H{=}100$, gains of up to $+17\%$ over VMD at the same $K$. The inference-time axis reinforces this advantage, with C-OMD remaining the fastest method.

In Fig.~\ref{fig:combined_results}(c) under channel impairment, C-OMD loses only ${\approx}16$ as SNR falls from $30$\,dB to $0$\,dB, while VMD sheds more than $34$ over the same range; the reason is structural, orthogonality confines noise injected into one frequency band to that mode alone, preventing it from cascading across the decomposition in a way that neither VMD nor SVMD can prevent. Taken together, the two panels establish a \emph{decomposition-first, attention-second} design principle: C-OMD provides the representational precondition for high accuracy, and cross-side attention yields an orthogonal, additive gain on top.

Each point in Fig.~\ref{fig:combined_results}(a) represents the average accuracy across 9 outputs corresponding to the three spatial dimensions (X, Y, Z) of the three features: F, V, and P. To provide a more detailed view of Fig.~\ref{fig:combined_results}(a), the average of each of the three features in three dimensions is provided in Table~\ref{tab:combined}. Table~\ref{tab:combined} delivers an unambiguous message: C-OMD+MDA is the Pareto-optimal choice across all measured axes. It achieves $0.089$ GFLOPs, representing a $2.5\text{--}3.6\times$ reduction compared to VMD ($0.23\text{--}0.32$) and a massive $19.1\text{--}22.4\times$ reduction compared to SVMD ($1.7\text{--}2.0$). Furthermore, with an inference time of approximately $0.06$\,ms, it provides an $\approx 8\times$ speedup over VMD and an $\approx 30\times$ speedup over SVMD. Most importantly, it maintains the highest accuracy across all window sizes, reaching $98.6\%$ at $H=1$ and $82.4\%$ at $H=100$. 
In short, C-OMD's explicit orthogonality simultaneously reduces computation and improves prediction; there is no accuracy--efficiency trade-off to navigate. 

\section{Conclusion}
In this work, we presented an orthogonal decomposition framework for decomposing a signal into its components. This allows for robust prediction in TI systems. By integrating C-OMD for mode extraction with the newly proposed MDA, we demonstrate that C-OMD+MDA consistently outperforms baselines, as orthogonality provides actionable insights that prevent interference between modes. Our framework lays the groundwork for adaptive TI systems that strike a balance between human control and semi-autonomous prediction.\\
\textbf{Limitations.} C-OMD needs the mode count $K$ fixed in advance, tuned per signal class, and we test it only on bilateral haptic traces from teleoperation.\\ 
\textbf{Future work.} We plan to extend to Discrete space and ECG signals, whose quasi-periodic, multi-component structure is well-suited to mode-domain prediction.
\bibliographystyle{IEEEtran}
\bibliography{refs}

\end{document}